\documentclass[11pt]{article}

\pdfoutput=1
\usepackage{moriond,epsfig}
\usepackage{amsmath}
\usepackage{graphicx}
\usepackage{verbatim}

\newcommand{\Bot}{$b$}
\newcommand{\Charm}{$c$}

\newcommand{\myW}{$W^{\pm}$}
\newcommand{\Z}{$Z$}

\newcommand{\ee}{$e^{+}e^{-}$}
\newcommand{\mumu}{$\mu^{+}\mu^{-}$}
\newcommand{\Zee}{$Z \rightarrow e^{+}e^{-}$}

\newcommand{\Wlnu}{$W \rightarrow \ell \nu $}
\newcommand{\Wenu}{$W \rightarrow e \nu $}

\newcommand{\ppbar}{$p\overline{p} $}
\newcommand{\ttbar}{$t\overline{t} $}

\newcommand{\Dzero}{D\O}
\newcommand{\alphas}{$\alpha_s$}

\newcommand{\ET}{$E_{T}$}
\newcommand{\pT}{$p_{T}$}

\newcommand{\MET}{$\not\!\! E_T $}

\newcommand{\abseta}{$\mid \eta \mid$}

\newcommand{\GeV}{GeV}
\newcommand{\GeVc}{GeV/$c$}

\newcommand{\invpb}{pb$^{-1}$}

\newcommand{\invfb}{fb$^{-1}$}

\newcommand{\plm}{$\pm$}

\newcommand{\gte}{$\ge $}

\def\Journal#1#2#3#4{{#1} {\bf #2}, #3 (#4)}

\def\PRL{\em Phys. Rev. Lett.}
\def\PRD{{\em Phys. Rev.} D}

\def\JHEP{{\em Jour. High En. Phys.}}

\def\be{\begin{equation}}
\def\ee{\end{equation}}
\def\bea{\begin{eqnarray}}
\def\eea{\end{eqnarray}}

\begin{document}
\vspace*{1cm}
\title{\myW / \Z\ + JETS AND \myW / \Z\ + HEAVY FLAVOR JETS AT THE TEVATRON}

\author{C. NEU} 

\address{Department of Physics and Astronomy, University of Pennsylvania			\\
Philadelphia, PA USA}

\maketitle\abstracts{Studies of the production of \myW /\Z\ + jets
are important for a variety of reasons.  Herein the latest Tevatron
results on these production mechanisms are reviewed with an emphasis
on comparison of data results to the latest theoretical models.}

\section{Motivation}

\myW /\Z\ + jets is a valuable sample for analysis at the Tevatron.
Its high statistics allow one to probe the validity of predictions
from perturbative Quantum Chromodynamics (pQCD).  These processes play
an important role in the Tevatron and LHC physics programs; \myW /\Z\
+ inclusive jets will be a valuable standard model calibration sample
at the LHC and \myW /\Z\ + heavy flavor are significant backgrounds to
top, Higgs and other new physics searches at both the Tevatron and
LHC.  State-of-the-art leading order (LO) and next-to-leading order
(NLO) calculations on these processes are the focus of several active
theory collaborations.  The predictions from these calculations would
benefit from experimental verification.

Below are described important Tevatron results on \myW /\Z\ + 
inclusive jets, \myW\ + single \Charm\ and \myW /\Z\ + \Bot -jets
and how these results compare to available theoretical predictions.

\section{\myW\ + Jets and \Z\ + Jets}

\begin{figure}[t!]
\vspace{-0.4cm}
$\begin{array}{c@{\hspace{-0.0in}}c}
\multicolumn{1}{l}{\mbox{\bf (a)}} &
\multicolumn{1}{l}{\mbox{\bf (b)}} \\ [0.0cm]
      \hspace{-0.2in}\scalebox{1.5}{\includegraphics[width=0.35\textwidth]{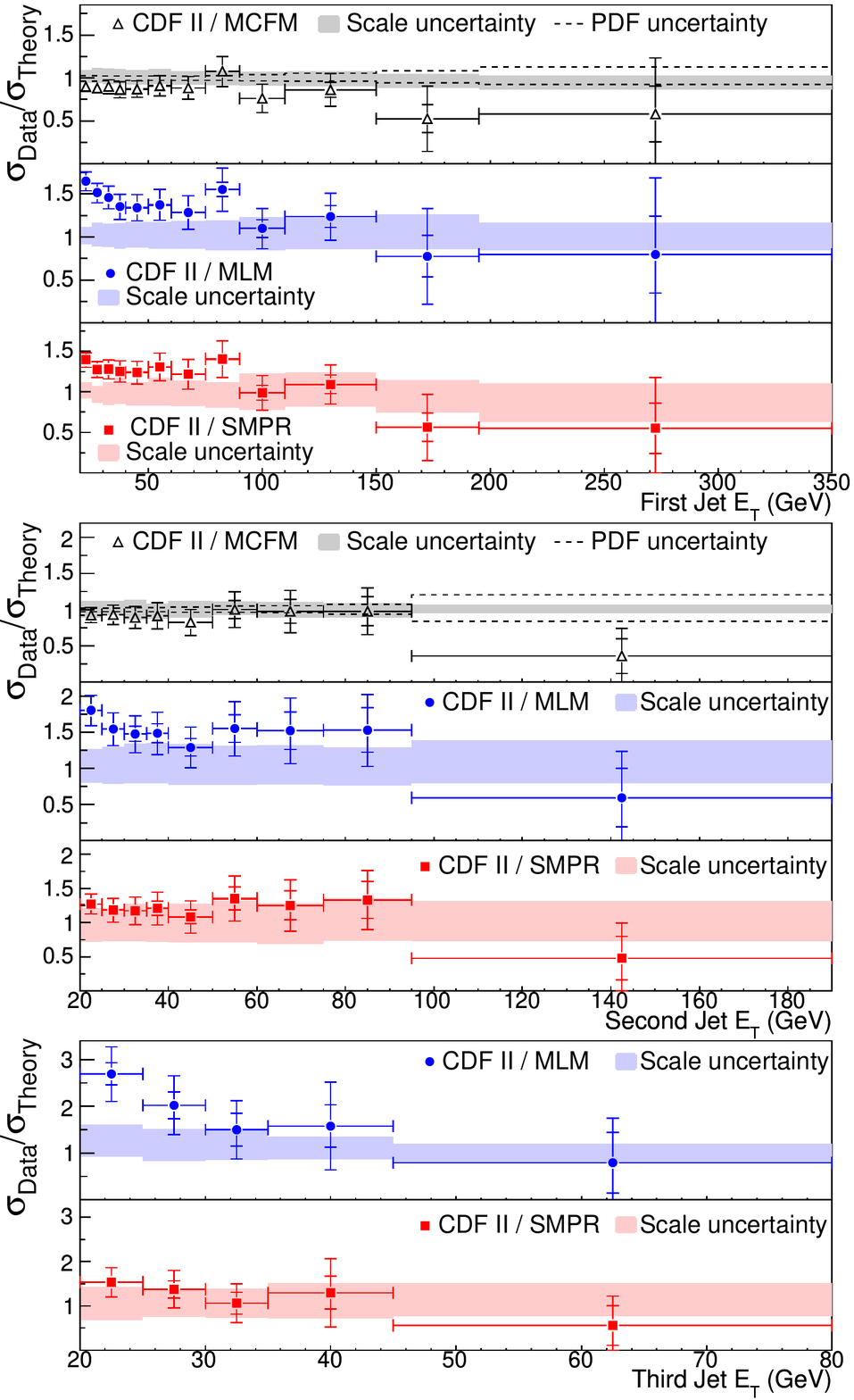}} &	
      \hspace{-0.2in}\scalebox{1.5}{\includegraphics[width=0.35\textwidth]{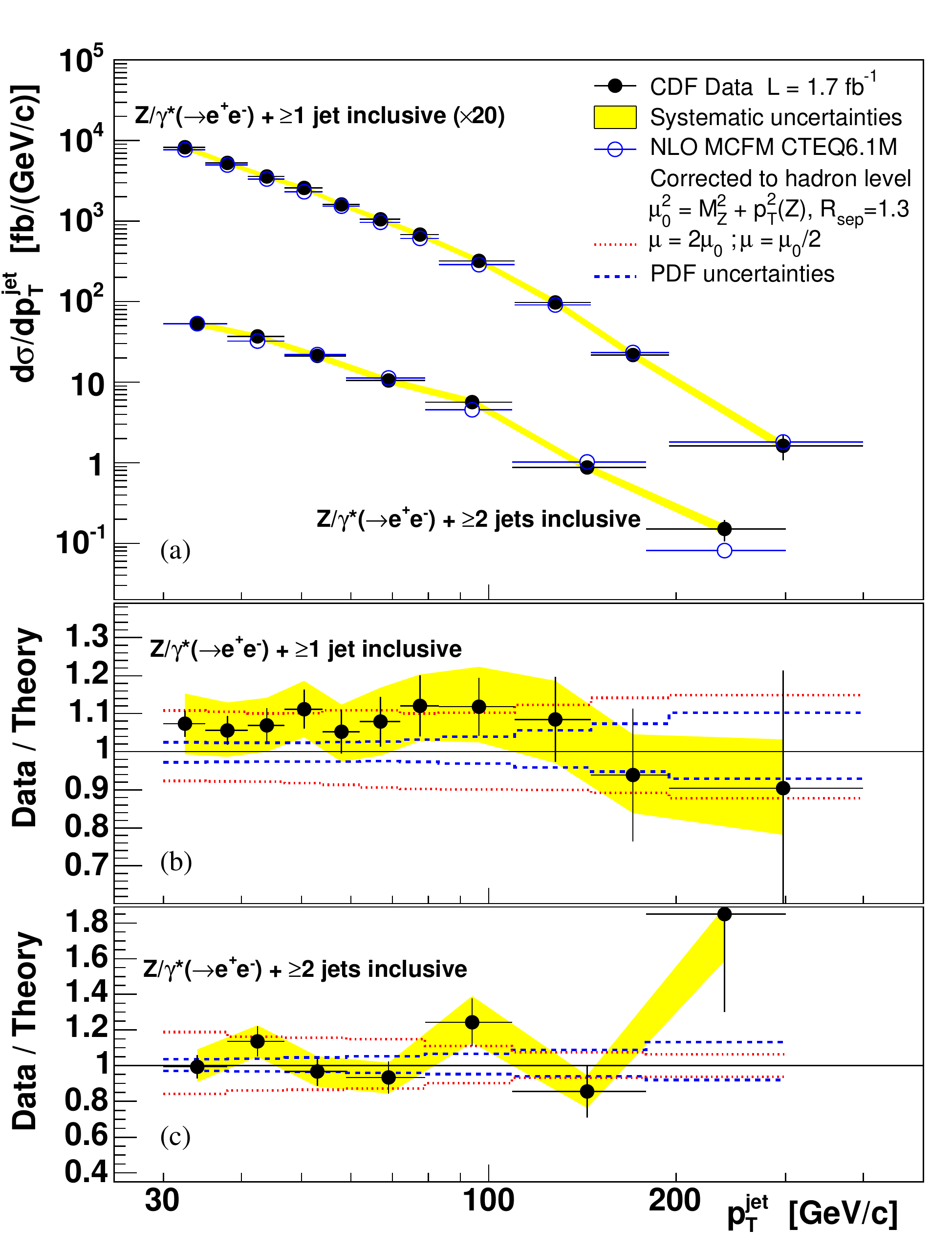}} \\ 
\end{array}$
\caption{Differential cross section comparison of data and several theoretical predictions for (a) first, second and third jet \ET\ in \myW\ + \gte\
1 jet events in 320 \invpb\ of CDF Run II data and (b) jet \pT\ in \Z\
+ \gte\ 1 jet and \Z\ + \gte\ 2 jets in 1.7 \invfb\ of CDF Run II
data.}\label{fig:wzjets}
\end{figure}

The CDF experiment has studied the production of jets in events with
\myW\ and \Z\ bosons~\cite{ref:wjets,ref:zjets}.  \Wenu\ events
are selected by identifying a high \ET , central electron along with
significant missing transverse energy, \MET ; \Zee\ events are
selected by requiring one such electron with another that is either
central or in the forward region of the calorimeter, with the invariant
mass of the electron pair required to be near the \Z\ mass peak.
Events are then assigned to bins of minimum jet multiplicity.  Major
sources of background in the \myW +jets analysis include events with
fake \myW 's and electroweak sources (\ttbar , single top, dibosons);
backgrounds in the \Z +jets analysis are dominated by multijet
production and \myW +jets events in which the
\Z\ signal is faked.  Acceptance for these events
is studied using simulated signal samples; the differential cross
section for the jets in these events is then examined and compared to
some available theory predictions as depicted in Figure~\ref{fig:wzjets}.

From Figure \ref{fig:wzjets}(a) one can see that the NLO prediction
from MCFM~\cite{ref:mcfm} is accurately reproducing the jet \ET\
spectrum in \myW + 1 or 2 jets.  For higher multiplicity events, LO
calculations are necessary.  The current preferred method for
generating such events at LO relies on generating multiple samples
using a matrix element calculation at fixed orders in \alphas\ and
then employing a parton shower program to add in additional soft,
colinear jets.  Matching algorithms have been designed to identify
events that could be double counted in this recipe.  From
Figure~\ref{fig:wzjets} (a) one can see that the LO
prediction consisting of the matrix element calculation from 
MadGraph~\cite{ref:madgraph}, parton shower from Pythia~\cite{ref:pythia} and
matching scheme from CKKW~\cite{ref:ckkw} is superior to that of
ALPGEN + Herwig shower + MLM-matching~\cite{ref:alpgen}.  It remains
to be understood which component of the prediction is causing the
difference in these LO predictions.  In Figure~\ref{fig:wzjets}(b) one
can see that the NLO prediction from MCFM accurately reproduces the
jet \pT\ spectrum in \Z +jets events, providing additional
confirmation of the validity of the NLO predictions.

\section{$W$+\Charm }

$W$+single-\Charm\ production is an important process at
the Tevatron.  $W$+\Charm\ offers insight on the strange content
inside the proton. The process also allows an opportunity to measure
$|V_{cs}|$ in a $Q^2$ regime not yet probed.  Also, $W$+\Charm\
contributes to the background to top production and prominent Higgs
search channels at the Tevatron.

CDF~\cite{ref:cdfwc} and \Dzero ~\cite{ref:dzerowc} have measured the $W$+\Charm\ process in Run II
using a similar strategy.  Leptonic $W$ decays (\Wlnu\ with $\ell =
e\textrm{ or }\mu$) are selected via a high \pT\ isolated central
lepton and large \MET .  Among the required jets in the selected
events, evidence is sought for semileptonic hadron decay through the
identification of a soft muon inside the jet cone.  It is a feature of
$W$+\Charm\ production that the electric charge of the $W$ and 
\Charm\ are opposite.  The sign of the \Charm\ quark is determined
from the charge of the muon used to identify semileptonic hadron decay.
An excess of opposite-sign primary lepton and soft muon events is
indicative of $W$+\Charm\ production.  Opposite sign backgrounds
include Drell-Yan production of \mumu , $Wq$ production and fake 
$W$'s.  

CDF measured in 1.7 \invfb\ of data the production cross section
for $W$+\Charm\ times the leptonic branching ratio of the $W$,
$\sigma(Wc) \times \mathrm{BR}(W \rightarrow \ell \nu )= $ 
9.8 \plm\ 2.8(stat) $^{+1.4}_{-1.6} $ (syst) \plm\ 0.6(lum) pb for 
events with $p_T^c >$ 20 \GeVc\ and \abseta\ $<$ 1.5.  This can be 
compared to the NLO prediction from MCFM of 11.0 $^{+1.4}_{-3.0} $.
\Dzero\ measured in 1 \invfb\ of data the ratio 
$R=\frac{\sigma(Wc)}{\sigma(W+\mathrm{jets})}$; measuring the
ratio has the virtue that numerous sources of systematic error
cancel out.  The result $R =$ 0.071 \plm\ 0.017 is reasonably
consistent with a LO prediction from ALPGEN of 0.040 \plm\ 0.003.

\section{\myW\ + \Bot -Jets and \Z\ + \Bot -Jets }

\begin{figure}[t!]
\vspace{-1in}
\centering
\includegraphics[width=0.5\textwidth]{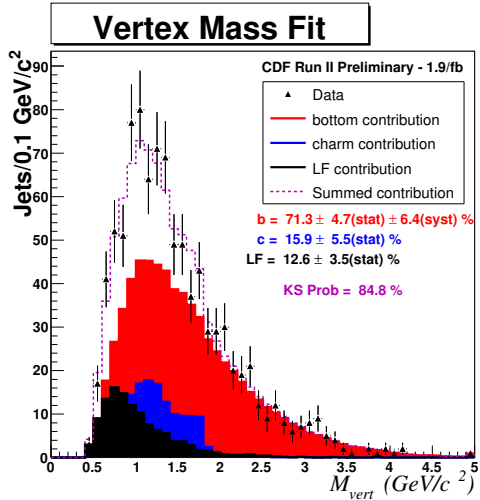}
\vspace{-0.8cm}
\caption{Vertex mass fit of tagged sample in CDF \myW\ + \Bot -jets analysis in 1.9 \invfb\ of data.
\label{fig:wbjets}}
\end{figure}

\myW /\Z +\Bot\ jet signatures are important backgrounds to top
and Higgs channels at the Tevatron.  Separate analyses were
undertaken to measure the \Bot -jet cross section in \myW\ and \Z\
events with increased precision in the hopes of improving the
understanding of these final states.

The event selection for the \myW +\Bot\ jets analysis is similar to
that employed in the $W$+\Charm\ analysis discussed above.  Here however
\Bot\ jets are selected via the identification of a secondary decay vertex
well-separated from the primary \ppbar\ interaction point.  Among the jets
possessing vertex tags, the \Bot\ content is extracted via a maximum
likelihood fit of the vertex mass, which is the invariant mass of the
charged particle tracks comprising the secondary vertex.  This variable
is discriminant among the different species of jets; from Figure~\ref{fig:wbjets}
one can see that among the tagged jets $\sim $ 71\% are found to be from
\Bot .  Backgrounds to this \myW +\Bot -jets signal include top production,
diboson production and fake \myW 's.  Signal acceptance was studied
with simulated \myW +\Bot -jet events using the ALPGEN event
generator.  Signal events are considered from a restricted region of
phase space ($e/\mu $ with \pT\ $>$ 20 \GeVc , \abseta\ $<$ 1.1, a
neutrino with \pT\ $>$ 25 \GeVc\ and exactly 1 or 2 \ET\ $>$ 20 \GeV ,
\abseta $<$ 2.0 jets) to avoid strong dependence on the signal model
in regions where we are not experimentally sensitive.

The \Bot -jet cross section in \myW\ events in 1.9 \invfb\ of CDF Run
II data was measured to be $\sigma_{b\mathrm{-jets}}(W+b\mathrm{-jets}) 
\times \mathrm{BR}(W\rightarrow \ell \nu ) = 2.74 \pm 0.27 (\mathrm{stat}) 
\pm 0.42 (\mathrm{syst}) \mathrm{pb}$, where
the systematic error is dominated by the uncertainty in the vertex
mass shape one assumes for \Bot\ jets.  This jet cross section result
can be compared to the prediction from ALPGEN of 0.78pb, a factor of
3-4 lower than what is observed in the data. Work is ongoing to
understand the difference.

\begin{figure}[t!]
\vspace{-0.0cm}
$\begin{array}{c@{\hspace{-0.0in}}c}
\multicolumn{1}{l}{\mbox{\bf (a)}} &
\multicolumn{1}{l}{\mbox{\bf (b)}} \\ [0.0cm]
      \hspace{-0.2in}\scalebox{1.5}{\includegraphics[width=0.35\textwidth]{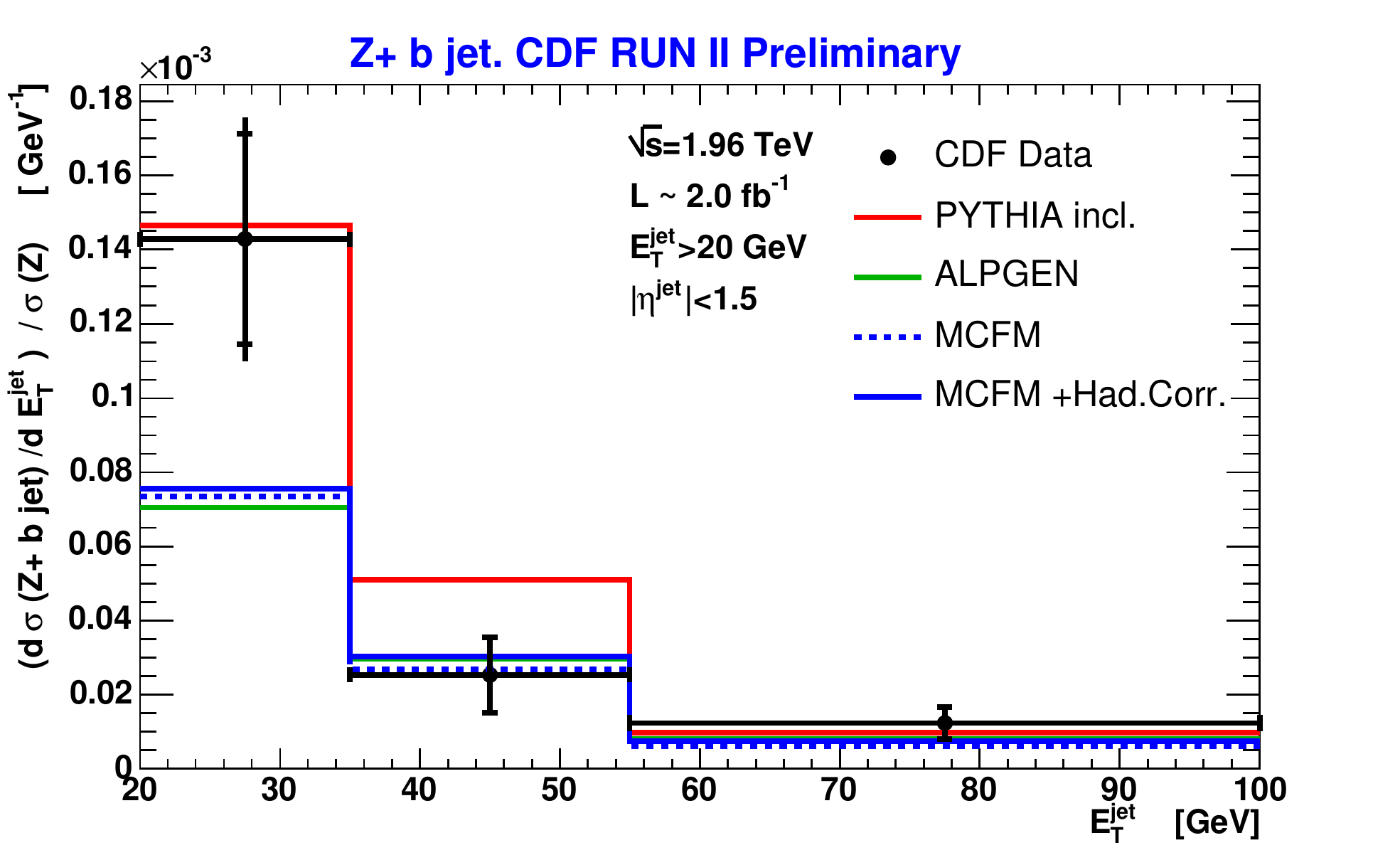}} &	
      \hspace{0.02in}\scalebox{1.5}{\includegraphics[width=0.35\textwidth]{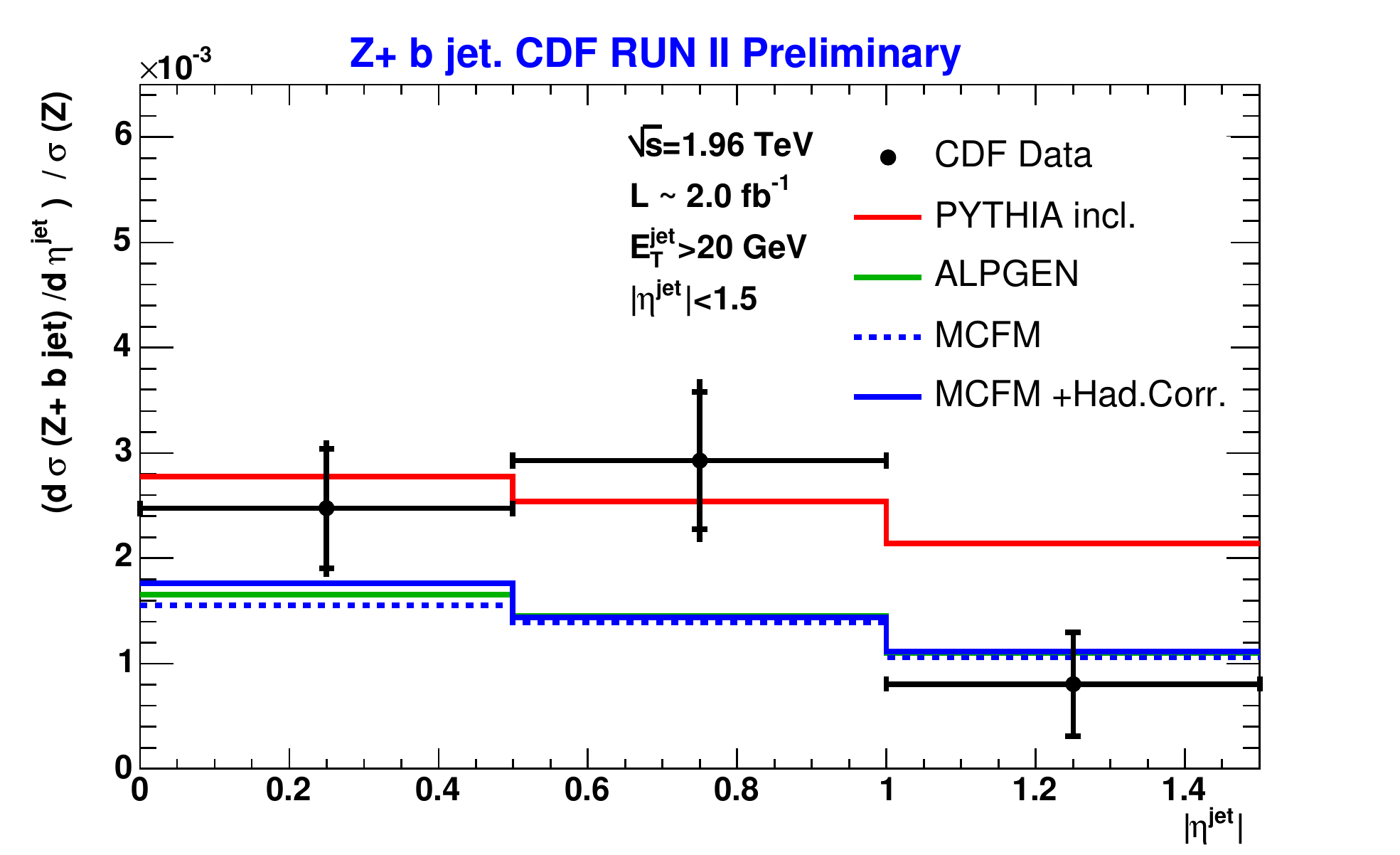}} \\ 
\end{array}$
\caption{\Z\ + \Bot\ jet differential cross sections as a function of jet \pT\ and \abseta\ from
CDF's 2 \invfb\ result.}\label{fig:zbjet}
\end{figure}

The \Z +\Bot -jet analysis used a similar technique to extract the
\Bot\ content of its tagged jet sample.  This analysis has succeeded
in examining differential cross sections for the \Bot\ jets in \Z\
events. From Figure ~\ref{fig:zbjet} one can see that the
differential \Bot -jet cross sections versus jet \pT\ (a) and \abseta\ (b) are
not reproduced in all bins by any of the predictions that were
constructed.  Pythia appears to do a reasonable job at low jet \pT\
but less so as the jet \pT\ increases.  The ALPGEN and MCFM
predictions are consistent with each other but not with the data
except for a few bins.  It remains to be understood why the
predictions are so different.

\section{Summary}

The \myW /\Z\ + jets samples at the Tevatron offer a valuable high
statistics testbed for state-of-the-art pQCD calculations. It appears
that for inclusive jet production the NLO predictions are accurately
describing the data for \myW /\Z\ + up to 2 jets.  Predictions for
higher parton multiplicity events at NLO would be beneficial.  As for
\myW /\Z\ + heavy flavor, NLO predictions for the integrated cross
section for \myW + single-\Charm\ appear to be accurate.  A consensus
on \myW /\Z\ + \Bot -jets has yet to be reached; both LO and NLO
predictions do not consistently reproduce the integrated or
differential rates of these events in the data.

\end{document}